\documentstyle[aps,epsf,twocolumn]{revtex}
\def\Lsu2{{\cal L}_{\mbox{SU(2)}}}

\def\su2xsu2{{SU(2)\times SU(2)}}
\def\su3xsu3{{SU(3)\times SU(3)}}

\def\ben{\begin{eqnarray}}
\def\een{\end{eqnarray}}

\newcommand{\be}{\begin{eqnarray}}
\newcommand{\ee}{\end{eqnarray}}

\newcommand{\beq}{\begin{equation}}
\newcommand{\eeq}{\end{equation}}
\newcommand{\bea}{\begin{eqnarray}}
\newcommand{\eea}{\end{eqnarray}}

\begin{document}

\draft
\title{\bf  QCD-Inspired Spectra from Blue's Functions}

\author{{\bf Maciej A.  Nowak}$^{1}$, {\bf G\'abor Papp}$^{2}$ 
and {\bf Ismail Zahed}$^3$}

\address{$^1$
GSI, Plankstr.1, D-64291 Darmstadt \& Institut f\"{u}r Kernphysik,
TH Darmstadt, Germany \& \\Department of Physics,
Jagellonian University, 30-059 Krakow, Poland;\\
$^2$GSI, Plankstr. 1, D-64291 Darmstadt, Germany \&\\
Institute for Theoretical Physics, E\"{o}tv\"{o}s University,
Budapest, Hungary\\
$^3$Department of Physics, SUNY, Stony Brook, New York 11794, USA.}
%\address{$^1$ Department of 
%Physics, Jagellonian University, 30-059 Krakow, Poland;\\
%$^4$Department of Physics, SUNY, Stony Brook, New York 11794, USA.}
\date{\today}
\maketitle

 \begin{abstract}
We use the law of addition in random matrix theory to analyze the spectral 
distributions of a variety of chiral random matrix 
models as inspired from QCD whether through symmetries or models. 
In terms of the Blue's functions recently discussed by Zee, we show that 
most of the spectral distributions 
in the macroscopic limit and the quenched approximation, follow algebraically 
from the discontinuity of a pertinent solution to a cubic (Cardano) or a 
quartic (Ferrari) equation. We use the end-point equation of the energy 
spectra in chiral random matrix models to argue for novel phase structures,
in which the Dirac density of states plays the role of an order parameter.

\end{abstract}
%\pacs{PACS numbers :  11.30.-j, 12.38.-t, 12.38.Aw, 12.90.+b.}

{\bf 1.\,\,\,}
Lattice simulations have established that QCD undergoes a 
phase transition at finite temperature \cite{LATTICE}. The precise  nature 
and character of the transition are still debated. The simulations, however, 
show that the QCD phase diagram with three flavors is rather involved, with a 
strong dependence on the value of the current masses. 

Important insights into the role played by chiral symmetry may be found in the
quark density of states at zero virtuality. Banks and Casher 
\cite{BANKS} have shown that at this point
the density of quark states is directly proportional 
to the quark condensate in the chiral limit. 
The spontaneous breaking of chiral symmetry is followed by a 
structural change in the quark eigenvalues near zero virtuality. 
It was also observed that some generic spectral oscillations take place 
in this region \cite{JAC,JUREK}, reminiscent of Airy oscillations 
\cite{BREZIN}.

An important question in QCD concerns the structure and character of 
the Dirac spectrum during a phase transition, as induced by changing 
the number of flavors, the character of the color representation,
the quark masses, the temperature or the baryon density, to cite a few.
Since chiral symmetry breaking is encoded in the eigenvalue distribution 
at zero virtuality, it is clear that a chiral phase transition would 
affect quantitatively this distribution. Since massive QCD does not have,
strictly speaking, a good order parameter, it is natural to study the behavior 
of the spectral density near zero virtuality as a potential alternative.
The possible change in the quark wavefunctions at zero virtuality is 
reminiscent of a phase transition from a delocalized and hence coherent phase, 
to a localized and hence incoherent phase. The coherent phase will be 
identified with the Goldstone phase in our case.

An efficient way to get  the Dirac spectrum  
has been established through the use of 
chiral random matrix models \cite{JAC,SHU}. Chiral random matrix models are 
0-dimensional field theories, inspired by the flavor and spin structure of  
QCD motivated Dirac operators. They share much in common with 
"two-level" Nambu-Jona-Lasinio models \cite{MACIEK}. By restricting the 
discussion to only constant modes, the role of symmetry and the lore of 
randomness become dominant and transparent. Results from random matrix
models both in the microscopic \cite{MICRO} and macroscopic \cite{JUREK}
limit have compared favorably with some lattice simulations of the QCD
spectra\cite{COLUMBIA,KALKREUTER}.

The purpose of this letter is to use newly 
developed methods by Zee \cite{ZEE} regarding the addition laws of random 
matrix theory, to discuss the generic character of 
the Dirac spectrum in the macroscopic limit and quenched approximation. 
For a large class of chiral random 
matrix models as inspired by QCD either in the continuum or on the lattice,
and instanton models, we show that
the resulting spectra follow simply from the general lore of random plus 
deterministic. In section~2, we introduce the concept of Blue's functions and 
review Zee's argument for the deterministic plus random (Gaussian) case. In 
section~3, we
show how the Blue's functions can be used to condition the 
end-points of the spectral distributions, resulting into quadratic equations
for positive end-points
(Cardano) or cubic equations for positive end-points (Ferrari). 
In section~4  we apply these results to massive Wilson fermions in
the one-flavor approximation, as well as massless QCD on the cylinder. The 
resulting spectra and singularities belong to the Cardano class. We show that
the spectral distribution of QCD with 2+1 flavors in the Wilson formulation,
follows from a linear 
superposition of the solutions to Cardano class. Instanton models with 
off-diagonal order fall into the Ferrari class.  Our conclusions and 
recommendations are summarized in section~5.

\vskip .5cm
{\bf 2.\,\,\,}
The spectrum of a single random matrix was determined long ago by Wigner
and others \cite{RANDOMOTHERS}. Generically, the averaged level 
distribution of an $N\times N$ random matrix $H$ (Hamiltonian) with a 
probability distribution 
\be 
P(H)=\frac{1}{Z} \exp [-N {\rm Tr} V(H)]
\label{first}
\ee
follows from the discontinuity of the resolvent
\be 
\nu(\lambda)=-\frac{1}{\pi}\lim_{\epsilon \rightarrow 0} 
{\rm Im} G(z=\lambda +i\epsilon)
   \label{spectral}
\ee
with
\be 
G(z)=\left< \frac{1}{N} {\rm Tr} \frac{1}{z-H} \right>
\label{greendef}
\ee
The average $<\ldots>$ is short for averaging with the weight~(\ref{first}).
For polynomial potentials the explicit form of (\ref{greendef})
is known\cite{BREZIN1}. In the simplest case of a quadratic polynomial,
the random ensemble is known as Gaussian with the Green's function
\be 
G(z)= \frac{1}{2}(z-i\sqrt{4-z^2})
\label{greengauss}
\ee
The discontinuity of (\ref{greengauss}) is just Wigner's semi-circle for 
Gaussian ensembles,
\be
\nu(\lambda)=\frac{1}{2\pi}\sqrt{4-\lambda^2} \,\,\,.
\label{semicircle}
\ee

The problem of adding random matrices can be reduced by using
Blue's functions, as recently discussed by Zee \cite{ZEE}. 
The Blue's  function $B(z)$ is just the functional inverse of the Green's 
function $G(z)$. Operationally
\be
B(G(z)) = G(B(z))=z
\label{blue}
\ee
The Blue's function for a constant matrix $c$ (random matrix of infinite 
weight) follows from the corresponding Green's function
\be
G(z)=\frac{1}{z-c}
\label{greencons}
\ee
through the substitution $z\rightarrow B(z)$ in (\ref{blue})
\be
B(z)=c+\frac{1}{z}
\label{constantblue}
\ee
Similarly, the Blue's function for the Gaussian ensemble with the Green's 
function (\ref{greengauss}) follows from (\ref{blue}) in the form
\be
B(z)=z+\frac{1}{z}
\label{gaussblue}
\ee

If $B_1 (z)$ and $B_2 (z)$ are the Blue's functions of two random matrices,
then the Blue's function $B_{1+2} (z)$
for the sum follows from the addition law \cite{ZEE}
\be
B_{1+2}(z)=B_1(z) +B_2(z) -\frac{1}{z}
\label{additionlaw}
\ee
Hence, the problem of finding the spectral distribution of the sum of two
random matrices boils down to the following algorithm :
First, construct the appropriate Green's functions $G_1$ and $G_2$.
Second, find their functional inverses, i.e. Blue's functions $B_1$ and 
$B_2$. Third, construct $B_{1+2}$ using (\ref{additionlaw}).
Fourth, functionally invert $B_{1+2}$ using the definition of the 
Blue's function, leading to $G_{1+2}$.
The spectral density of the sum is the discontinuity of $G_{1+2}$ along the 
real axis.

To illustrate these points, consider with Zee \cite{ZEE} the case where $H$ 
is the sum of a deterministic matrix $H_D$ with eigenvalues $\epsilon_i$, 
($i=1,\ldots,N$) and a random Gaussian matrix $H_R$. 
The Green's function for the deterministic matrix  $H_D$ reads
\be 
G_D(z)=\frac{1}{N}\sum_i \frac{1}{z-\epsilon_i}
\label{greendeter}
\ee
Through the replacement $z\rightarrow B(z)$ the Blue's function $B_D$ is
\be
z=\frac{1}{N}\sum_i \frac{1}{B_D(z)-\epsilon_i}
\label{help1}
\ee
The addition rule gives
\be
B(z)=B_D(z) + B_R(z) -\frac{1}{z}= B_D(z) +z
\label{help2}
\ee
where the second equality comes after we have used the known form of the Blue's
function for the random Gaussian ensemble (\ref{gaussblue}).
Substituting $z\rightarrow G(z)$ in (\ref{help1}) and (\ref{help2})
we get a pair of equations
\be
G(z)&=& \frac{1}{N}\sum_i \frac{1}{B_D(G(z))-\epsilon_i}\nonumber\\
\nonumber \\
z &=& B_D(G(z))+G(z)
\label{help3}
\ee
from which we eliminate $B_D(G)$\footnote{Note that the explicit form of 
$B_D(z)$ is actually not needed.} to get
\be
G(z)=\frac{1}{N}\sum_i^N \frac{1}{z-G(z)-\epsilon_i}
\label{Pastureq}
\ee
This result was first established by Pastur\cite{PASTUR}. Zee's argument 
adds to its transparency. We also note that (\ref{Pastureq}) is nothing
but the resummed rainbow diagrams \cite{BREZINZEE}. It has the suggestive
form of a gap equation
\be
G(z)=\frac{1}{N}{\rm Tr} \left( \frac{1}{z-G(z)- H_D}\right)
\label{gap}
\ee
familiar from mean-field (large $N$) approaches. In 0-dimensional field 
theories and for a Gaussian measure, the self-energy reduces to the two-point
function. Pastur's equation is 
a polynomial in $G$ of degree $p+1$. Algebraic solutions are only available
for $p\leq 3$. Below we will show that a variety of QCD inspired random matrix 
models fulfills this restriction, with a chiral condensate in the massless case
given by
\be 
<\bar{q} q> = {\rm Im} \,\,G(z = i\epsilon )\,\,\,.
\label{bankscasher}
\ee

\vskip .5cm
{\bf 3.\,\,\,}
The knowledge of the Blue's functions allows for a direct assessment 
of the structural character of the spectral distribution (energy bands). 
This is particularly important for systems undergoing a phase change.
In general, a given spectrum has a support $L$ defined by the sequence 
of branch points
$\pm A_i, i=1,\ldots,p$  with positive real $A_i$ and $A_1>A_2>\ldots
>A_p$ consisting of  $p$ arcs, $[-2A_1,-2A_2],\ldots,[2A_2,2A_1]$ 
on which the spectrum is non-zero. Since the density of  states 
at the end-point behaves like $\sqrt{A-\lambda}$, its location
is determined by the condition
\be
\frac{dG}{dz}|_{z=A}  = \infty
\label{endpoint}
\ee
This condition can be translated into a condition on  the Blue's function 
\cite{ZEE}.
Let $F(z, B(z))$ be Pastur's  equation (\ref{Pastureq}) rewritten
in terms of the Blue's function, i.e.
\be
F(z, B(z)) \equiv  \frac{1}{N}\sum_i \frac{1}{B(z)-\epsilon_i-z} - z = 0
\label{blueeq}
\ee
%Substitute $A$ to $B(z)$ in (\ref{blueeq}), and assume that $A$ is 
%z-independent. 
Then the conditions determining the end-points of the 
spectral distribution follow from the simultaneous solutions to
\be
&&\frac{dF(z,B(z))}{d z} = 0 \nonumber \\
\nonumber \\
&&F(z,B(z))=0
\label{Zeepairs}
\ee 
with 
\be
\frac{dB(z)}{dz}|_A=0
\label{primzero}
\ee
The latter condition is just the end-point condition (\ref{endpoint})
rewritten in terms of the Blue's function.
The set of all end-points $\pm A_i$ in the spectral distribution
comes from solving for $B$ in the above set of equations
(\ref{Zeepairs},\ref{primzero}).

When Pastur's equation is cubic in $G$ (Cardano class), the corresponding 
Blue's equation (\ref{blueeq}) is quadratic, and the end-point conditions 
(\ref{Zeepairs}) follow from the solutions to a linear and biquadratic 
(quartic) equation, resulting into four roots
$\pm A_1, \pm A_2$. Therefore the spectral distribution consists of two arcs.
The condition when $A_2$ vanishes determines a transition point, with the
spectral distribution playing the role of an order parameter. Cardano class
consists of two phases $P_1$ and $P_2$, and the spectral distribution is 
either supported by one arc ($P_1$-phase) or two disconnected arcs 
($P_2$-phase).

When Pastur's equation is quartic in $G$ (Ferrari class),
the corresponding Blue's equation (\ref{blueeq}) is cubic,
and the boundary conditions reduce to the simultaneous solutions
of a quadratic and cubic equation, thereby maximally
six roots $\pm A_1, \pm A_2, \pm A_3$. Therefore the support for the spectral 
distribution consists of either three, two or one arcs, corresponding 
to the phases $P_3$, $P_2$, or  $P_1$. The critical points are determined 
from the condition that the real positive points $A_3$ and/or $A_2$ vanish.
We note that the present analysis can be easily extended numerically to 
higher degree polynomials, in the absence of the explicit Green's functions. 
This is particularly important for multiflavor QCD.

\vskip .5cm
{\bf 4.\,\,\,\,}
As a direct application of the ideas discussed above, we now consider four
different chiral random matrix models as inspired from the symmetries of the 
QCD Dirac operator and also instanton calculations.

\begin{figure}
\centerline{\epsfxsize=7cm \epsfbox{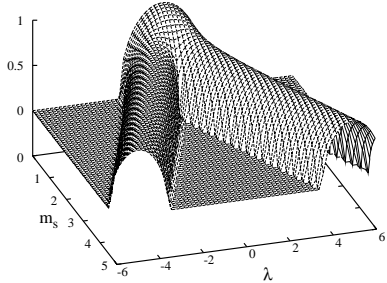}}
\caption{\label{fig1} Spectral distribution for one flavor (Wilson fermions)
as a function of the eigenvalue $\lambda$  and the quark mass $m_s$.}
\end{figure}

\vskip .25cm
$\bullet$  Lattice with 1 flavor.
\vskip .12cm
Recently, Kalkreuter\cite{KALKREUTER} has performed 
detailed lattice analysis of 
the distribution of eigenvalues for the (hermitean) Dirac operator
${\bf Q}_1=\gamma_5(D\!\!\!\!/ +m)$ using Wilson fermions in the quenched
approximation. The multiplication by 
$\gamma_5$ implies that ${\bf Q}_1^{\dagger} = {\bf Q}_1$. In terms of random 
matrix ensembles, Wilson fermions are represented by
the Gaussian Unitary Ensemble (GUE) for $N_c > 2$ and
by the Gaussian Orthogonal Ensemble (GOE) for $N_c=2$. For one-flavor, we have
\be 
{\bf Q}_1  = \left( \begin{array}{cc} m & 0\\
                                  0 &-m \end{array}\right)
+ {\bf R} \,\,\,\,.
\label{kalksum}
\ee
All the entries in the deterministic matrix are $N\times N$ valued,
while the random and hermitean matrix ${\bf R}$ is $2N\times 2N$
valued. Because the Wilson r-terms (needed to eliminate the quark doublers)
break explicitly chiral symmetry, ${\bf R}$ is not block-off diagonal. 
So while the eigenvalues of ${\bf R}$ are not paired, 
the eigenvalue distribution associated to ${\bf R}$ is symmetric about zero 
virtuality. Randomness populates evenly the positive and 
negative states.
For one flavor and in the massless case, the spectrum is hoped to be  
$U(1)_L\times U(1)_R$ symmetric on the average. The Green's function of the 
random matrix in 
(\ref{kalksum}) for the Gaussian ensemble is given by (\ref{greengauss}).
The resolvent $G(z)$ for the deterministic and random problem (\ref{kalksum})
is one of the solution   to the equation\cite{JUREK}
\be
G^3-2zG^2 +(z^2-m^2+1)G-z=0
\label{cubic1}
\ee
Kalkreuter's lattice spectra for one Wilson fermion belong to the Cardano 
class. A typical spectral distribution following from (\ref{cubic1}) 
is shown in Fig.~\ref{fig1}. It yields a 
structural change (transition from $P_1$ to $P_2$) for a critical  mass $m_*=1$,
with a density of states at zero virtuality $\nu (\lambda ) \sim \lambda^{1/3}$
\cite{JUREK}. The localization of the heavy quark, suggests that the flavor
symmetry is broken to $U_V (1)$ with the appearance of a heavy quark symmetry,
$i.e.$ invariance under spin-flip. It should be clear, however, that in our 
random analysis, (\ref{kalksum}) does not account for the $U_A(1)$ anomaly. 
The resulting spectral distribution, however, is qualitativaly similar to the
one obtained by Kalkreuter on the lattice using Wilson fermions \cite{JUREK}.

In physical units $m_*= 1/\Sigma =100-200$ MeV, which is rather 
close to the  strange quark mass in QCD. The origin of the scale $\Sigma$
follows from the Banks-Casher relation. Indeed, for two flavors
$<\overline{q} q > = -2\Sigma N/V_4$ \cite{BANKS}, where the chiral 
condensate is $<\overline q q > = <\overline{u}u +\overline{d}d>$,
with typically $<\overline{u}u> \approx <\overline{d}d> =
-((200-250) \,\,{\rm MeV})^{3}$. The density of quark zero modes is 
roughly $ n= N/V_4 \sim 1$ fm$^{-4}$. 

\vskip .25cm
$\bullet$ Finite Temperature.
\vskip .12cm
Recently, Jackson and Verbaarschot\cite{JACKSON}  have suggested a simple 
chiral random matrix model for the massless QCD Dirac operator 
$iD\!\!\!\!/$
on the cylinder, namely
\be 
{\bf Q}_2  = \left( \begin{array}{cc} 0 & \pi T \\
                                \pi T   &0 \end{array} \right)
+ \left( \begin{array}{cc} 0 & R \\
                                  R^{\dagger} &0 \end{array} \right)
\label{jacsum}
\ee
where $\pi T$ is short for one of the two lowest Matsubara frequencies. 
Again all entries in (\ref{jacsum}) are $N\times N$ matrix valued. 
Although the low 
temperature phase involves many Matsubara frequencies, (\ref{jacsum}) 
may be viewed as a schematic truncation with the right zero and high 
temperature content \cite{MACIEK}. 

This model belongs to the Cardano class. Indeed, 
the Green's function for the deterministic (thermal part) could be 
easily rewritten in a diagonal form by squaring the operator, 
with the result  
\be
G_T (z) = \frac{z}{z^2- (\pi T)^2}
\label{thermalgreen}
\ee
Adding the corresponding Blue's function we recover the cubic equation
(\ref{cubic1}) with the substitution $m\rightarrow \pi T$. This result
is expected from dimensional reduction arguments \cite{ZAHEDHANS}. At high 
temperature, and after suitable chiral rotations, each Matsubara mode carries
 a thermal mass, the lowest being  ${\bf m} (T) =\sqrt{m^2 +\pi^2 T^2}$.
The corresponding spectral distribution has been discussed numerically 
\cite{JACKSON}, and analytically \cite{STEPHANOV}.
The critical temperature is $T_* =1/\pi$. In dimensionfull
units it is $T_* = 30 - 70 $ MeV. Of course, this is
 rather unreasonable,  but so 
is (\ref{jacsum}). The addition of more Matsubara modes affects quantitatively 
$T_*$ \cite{MACIEK}. 

The critical temperature may also be assessed analytically by analyzing
the end-points of the spectrum. Indeed, in the Cardano class the end-points
$\pm A_1, \pm A_2$ for the thermal problem are located at
\be
A_1 &=& \frac{1}{2^{3/2}\pi T} 
\frac{(4\pi^2 T^2 -1 +\sqrt{8\pi^2 T^2 +1})^{3/2}}{\sqrt{8\pi^2 T^2 +1}-1}
\nonumber \\
A_2 &=& \frac{1}{2^{3/2}\pi T} 
\frac{(4\pi^2 T^2 -1 -\sqrt{8\pi^2 T^2 +1})^{3/2}}{\sqrt{8\pi^2 T^2 +1}+1}
\label{hotpoints}
\ee
The same equations govern the evolution of the end points in the one-flavor
case discussed above and in \cite{JUREK}, with the identification $m=\pi T$,
as well as the end-points of the  energy bands in the case of spin-dependent 
scattering off impurities in quantum Hall fluids \cite{ZEE}.
The real part of $A_2$ vanishes at the critical point $T_* =1/\pi$.

\vskip .25cm
$\bullet$ Lattice with $2+1$ flavors.
\vskip .12cm

The issue of chiral symmetry breaking in QCD is particularly interesting for 
the case of two light (up and down)  and one heavy  (strange) flavors. On the
lattice this is usually difficult to achieve with staggered fermions. It is 
best approached using Wilson fermions. 
Consider the case where the up and down quark mass are small but the
strange quark mass is kept fixed (2+1 flavors). Now, 
let us consider with Kalkreuter the spectral distribution for
the hermitean Dirac operator ${\bf Q}_3 = \gamma_5 (D\!\!\!\!/ + M)$ with 
$M={\rm diag} (m_u, m_d ,m_s)$ (with $m_u=m_d \approx 0$) and Wilson fermions. 

\begin{figure}
\centerline{\epsfxsize=7cm \epsfbox{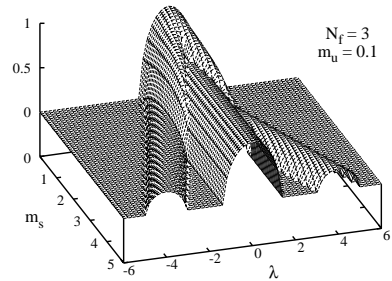}}
\caption{\label{figextra} Spectral distribution for 2+1 flavors
(Wilson fermions), as a function of the eigenvalue $\lambda$  and 
the strange quark mass $m_s$.}
\end{figure}

In this case, the Green's function is given by
\be
G(z)=\left< \frac{1}{6N} {\rm Tr}\, \frac{1}{z - {\bf Q}_3}\right>
\label{err1}
\ee
with 
\be
{\bf Q}_3=\bigotimes_{i=u,d,s} \left(\left(\begin{array}{cc} m_i & 0\\ 0 & 
-m_i\end{array} \right) + {\bf R} \right)
\label{latexart}
\ee
Each entry in the deterministic matrix is again $N\times N$ valued, while 
the random 
matrix is hermitean and $2N\times 2N$ valued. Due to the block-diagonal
structure of ${\bf Q}_3$, the resolvent (\ref{err1}) is the sum of three 
resolvents, each of the Cardano type. Specifically,
\be
G=\frac{2}{3}G_{m_u} + \frac{1}{3}G_{m_s}
\label{err2}
\ee
where $G_m$ are solutions to (\ref{cubic1}) with the corresponding masses. 
Figure~\ref{figextra} shows the corresponding spectral function 
for two light quarks $m_u=m_d=0.1$, as a function of the heavy quark mass 
$m_s$.

For small values of $m_s$, the spectrum is $U(3)_L\times 
U(3)_R$ symmetric, with soft breaking by the up, down and strange
quark mass. (The assumption being that the r-terms do not affect
considerably the chiral structure in the Wilson formulation).
For large values of $m_s$, the spectrum is $U(2)_L\times U_R(2)$
with an additional invariance under a spin-flip of the strange quark
(heavy-quark symmetry). We recall that the critical quark mass for the Cardano
class is $m_* =1$ or $100-200$ MeV in physical units. For $m_s >m_*$,
the spectral distribution resembles the $P_3$ phase expected from a quartic 
equation (Ferrari). However, the two are distinguishable by the evolution 
of their end-points.

\vskip .25cm

$\bullet$ Instanton Gas-Liquid
\vskip .12cm
In the instanton model of the QCD vacuum, the interaction between the 
instantons can cause them to cluster or dissociate into a gas, depending on 
the density of instantons per unit volume. Early simulations using instantons 
have revealed an admixture of free instantons (gas) and instanton clusters
at low instanton density\footnote{We note that this 
is the same as the density of quark zero 
modes, because of the Atiyah-Singer index theorem.}. At a critical density
$n_*\sim 1$ fm$^{-4}$ a 
random phase (intermediate between liquid and gas) is formed, that breaks 
spontaneously chiral symmetry \cite{EARLY}. Similar ideas have also been 
used to mock up the chiral transition in the instanton vacuum at finite 
temperature, using the Euclidean temporal direction as a mean of 
polarization \cite{SHURYAKMOL}. For these problems, the generic structure 
of the Dirac operator truncated to the space of zero modes is
\be
{\bf Q}_4 = \left( \begin{array}{cc} 0 & D \\
                                D   &0 \end{array}\right)
+ \left( \begin{array}{cc} 0 & R \\
                                  R^{\dagger} &0 \end{array}\right)
\label{weidensum}
\ee
The matrix $D$ carries off-diagonal short (molecule) or medium (cluster) range 
order. It is $N\times N$ valued. The matrix elements of $D$ in instanton 
models are identified with the hopping matrix elements  between instantons
and antiinstantons, hence chirality odd. 
For simplicity, we will restrict our discussion to the case where $D$ is
diagonal since
a molecule only involves one instanton and one anti-instanton at a time. 
The elements of the matrix $D$, however, need not be equal.

\begin{figure}
\centerline{\epsfxsize=7cm \epsfbox{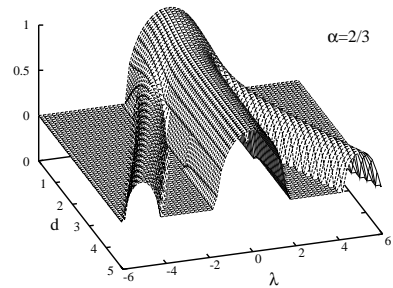}}
\caption{\label{fig2}Spectral distribution for the instanton model
(gas-liquid) as a function of the eigenvalue $\lambda$  and the hopping
strength $d$, with fixed $\alpha=0.66$.}
\end{figure}

Equation (\ref{weidensum})
is the sum of a deterministic $D$ and a random $R$ matrix. The case where 
$D = {\rm diag} ( d, d, ..., d)$ yields to the Cardano class discussed above. 
The case where some fraction $\alpha$ of the elements along the diagonal of $D$ 
are zero, and all the others are non-zero and equal to $d$ yields a Green's 
function for the molecules of the form
\be
G_{mol} (z) =\alpha \frac{1}{z} + (1-\alpha)\frac{z}{z^2-d^2}
\label{molgreen}
\ee
The addition law of the Blue's functions results in a quartic equation. 
The spectrum follows from the discontinuity of the pertinent  solution to
\be
G=\alpha \frac{1}{z-G} + (1-\alpha)\frac{z-G}{(z-G)^2-d^2}
\label{quartic2}
\ee
 The problem is of the Ferrari class.  The roots of (\ref{quartic2}) are known 
algebraically, so they can be used to derive explicitly the spectral 
distribution as shown in Fig.~\ref{fig2}.

The end-point singularities of (\ref{quartic2}) are given by 
\be
A_i = \pm \frac{x_i^2 +x_i(1-d^2) -\alpha d^2}{\sqrt{x_i}(x_i-d^2)}
\label{cubicband}
\ee
where $x_i$\,\,($i=1,2,3$) are the real and positive roots of the cubic 
equation
\be
x^3-x^2(1+2d^2)+xd^2(d^2+3\alpha-1)-\alpha d^4=0
\label{cubiceq}
\ee
For $\alpha =2/3$, Figure~\ref{fig2} shows a structural change in the 
spectral distribution for a hopping strength $d*\approx \sqrt{3}$ or in 
physical units $d_*= 170-340$ MeV.

\begin{figure}
\centerline{\epsfxsize=7cm \epsfbox{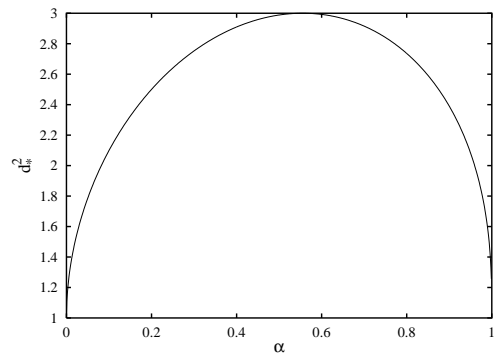}}
\caption{\label{fig0}Behavior of the hopping strength  $d$ 
as a function of $\alpha$.}
\end{figure}

 The critical value $d_*$ depends on 
the concentration of molecules ($1-\alpha$) as shown in Fig.~\ref{fig0}.
The structural change occurs away from zero virtuality.

In the vicinity of zero, (\ref{quartic2}) reduces to 
\be
G^4(0)+(1-d^2)G^2(0) -\alpha d^2=0
\label{weiden2}
\ee
leading to the condensate (in dimensionless units)
\be
|<\bar{q}q>| =  \frac {\pi}{\sqrt 2}\sqrt{1-d^2 +\sqrt{(1-d^2)^2 +4\alpha d^2}}
\label{weidcond}
\ee
The sign of the condensate depends on the way we approach zero virtuality in 
the complex $z$-plane. This result is similar to the one discussed by Wettig, 
Sch\"{a}fer and Weidenm\"{u}ller \cite{WEIDENMULLER} using different 
arguments\footnote{
The parameter $\alpha$ used in \cite{WEIDENMULLER} corresponds to our
$(1-\alpha)$.}.

\begin{figure}
\centerline{\epsfxsize=7cm \epsfbox{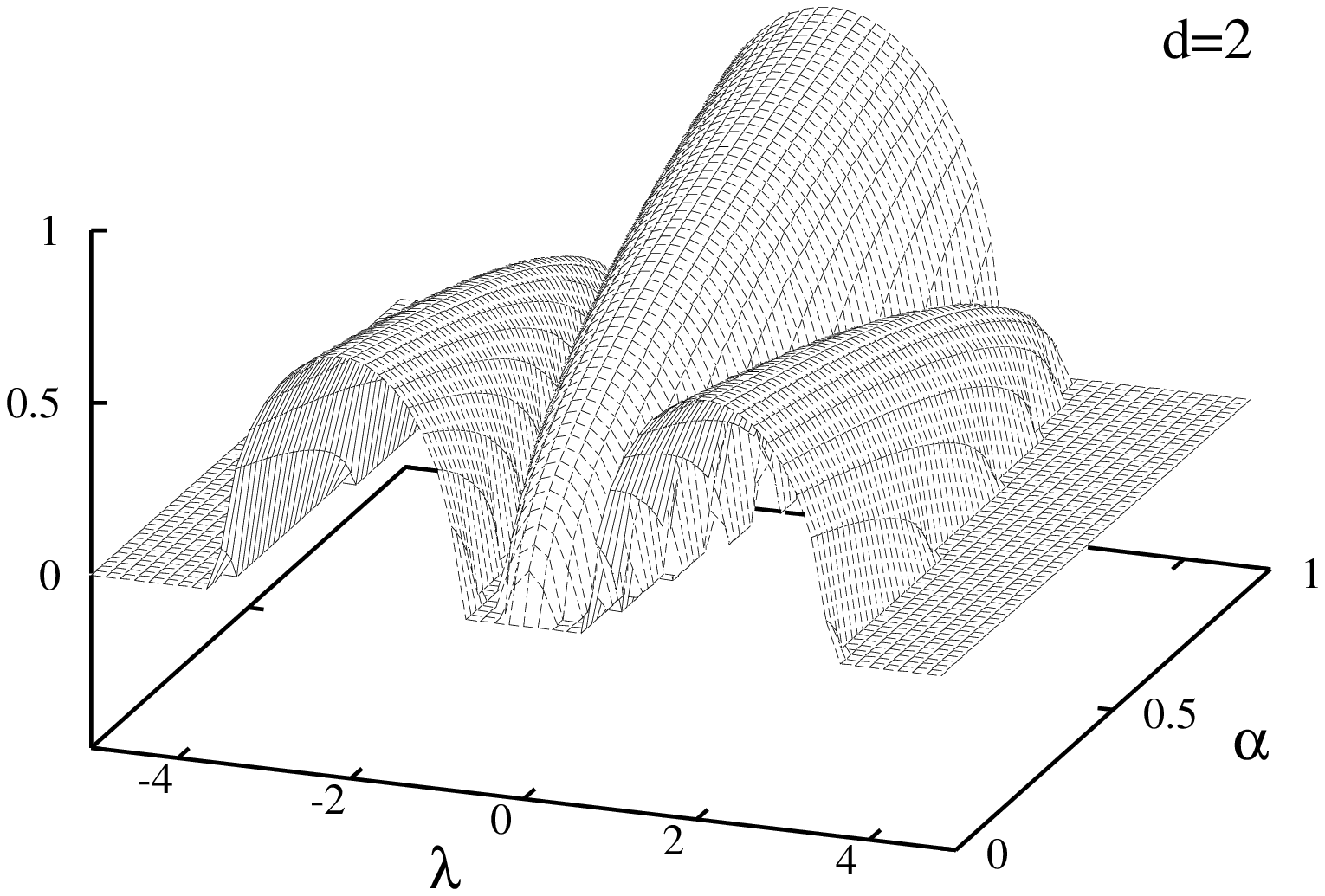}}
\centerline{\epsfxsize=7cm \epsfbox{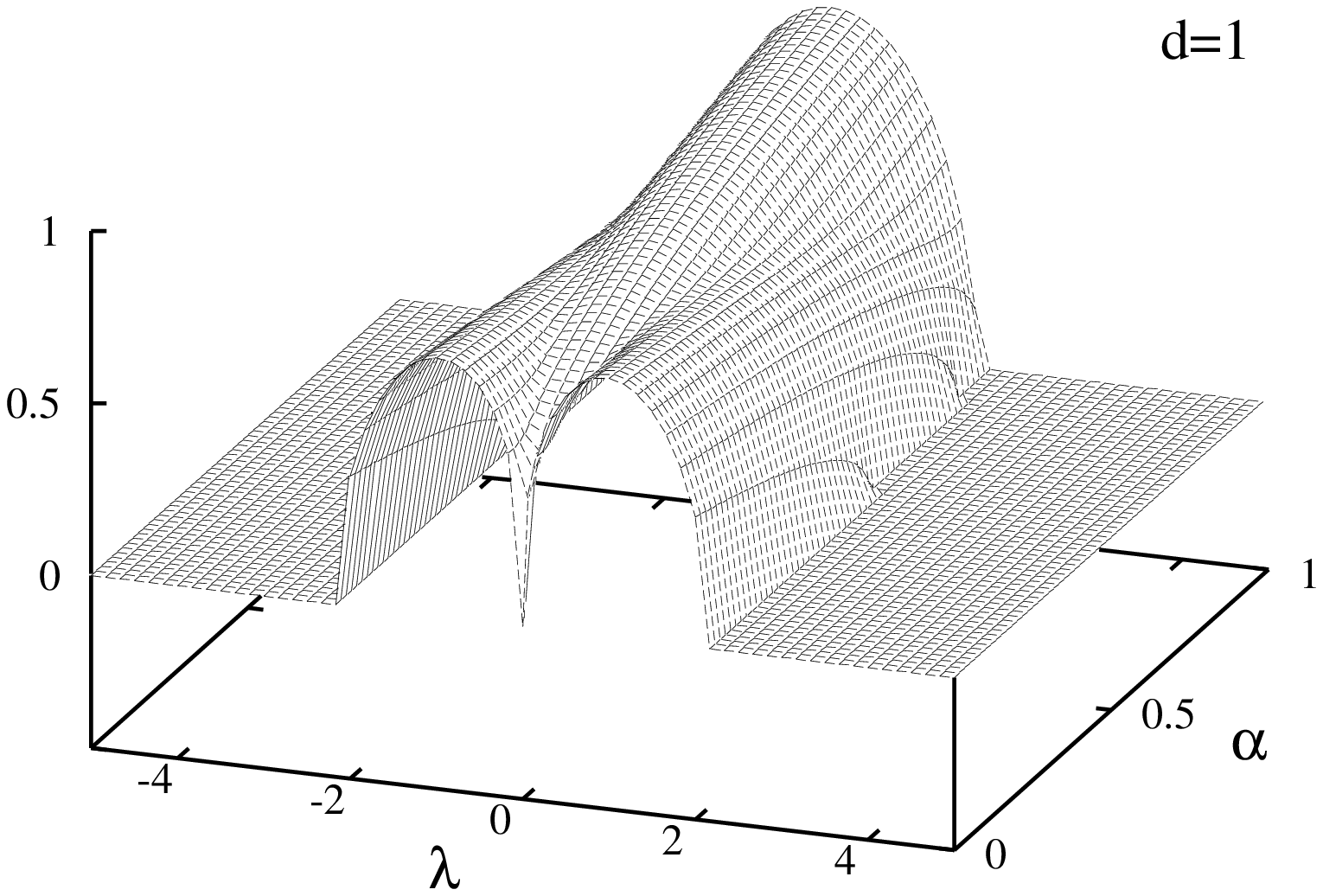}}
\centerline{\epsfxsize=7cm \epsfbox{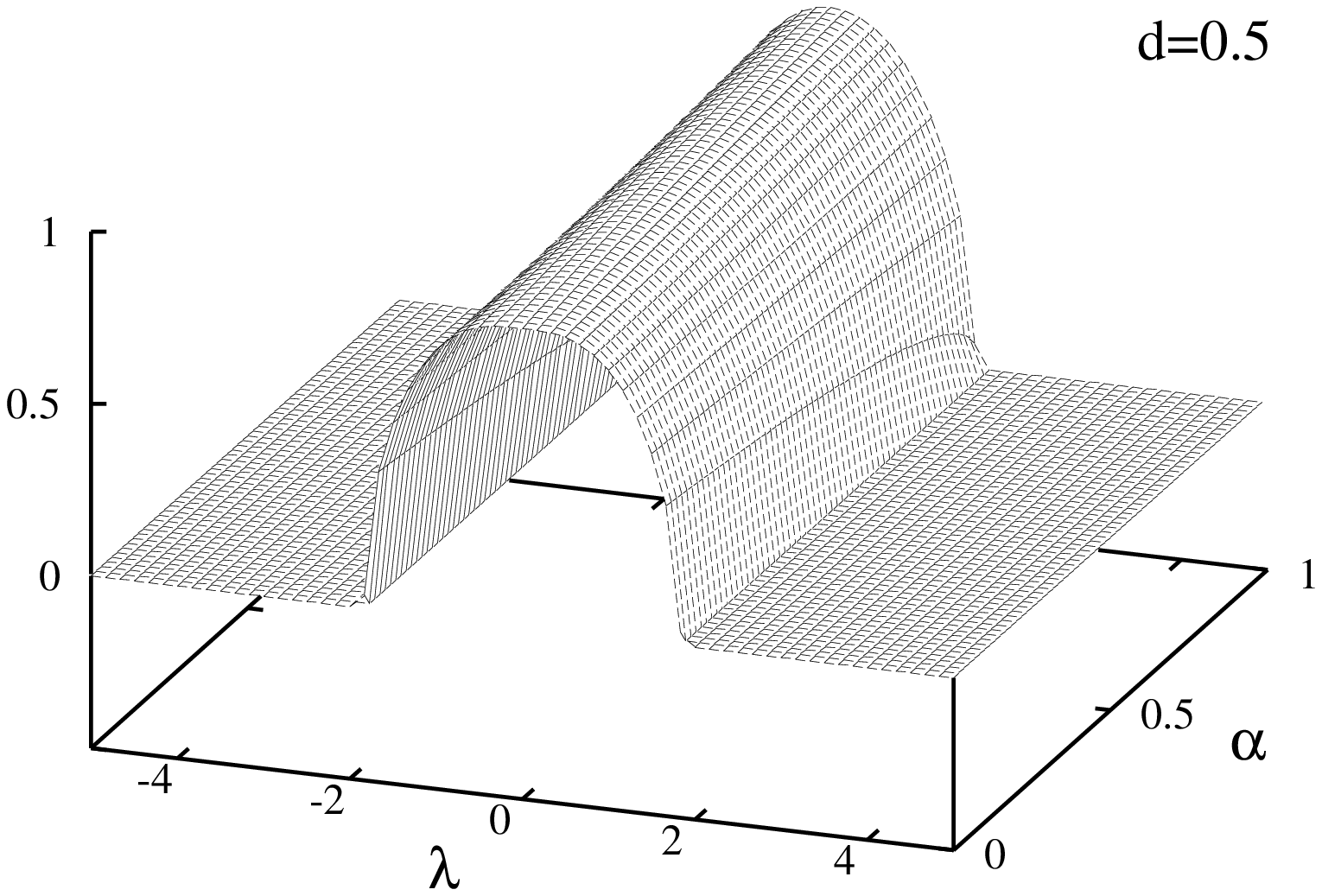}}
\caption{\label{fig3a}Spectral distributions for the instanton model
as a function of the eigenvalue $\lambda$, the concentration  $\alpha$ and 
the hopping strength $d=2$ (upper), $d =1$ (middle), and $d = 0.5$ (lower).}
\end{figure}

The critical points of (\ref{quartic2}) follow from the behavior of the 
end-points given  by (\ref{cubicband}-\ref{cubiceq}).
 In terms of the hopping strength $d$, the
spectral distributions shown in Figs.~\ref{fig3a},
 display novel phase structures as a function of the concentration $\alpha$. 
The upper figure  corresponds to the case where $d$
is greater than the maximal critical value $d_*=\sqrt{3}$. For 
$\alpha =1$ we are in the 
totally random phase with Wigner's semi-circle. There are no molecules. 
This is the $P_1$ phase.
For $0 <\alpha <1$ we have cohabitation phases ($P_3$ case) 
between a chirally symmetric
molecular phase and a spontaneously broken  chiral phase with a 
non-vanishing chiral condensate. For $\alpha=0$ all the elements along the 
diagonal of $D$ are equal to $d$. In this case the spectral 
function reduces to the Cardano class, see Fig.~\ref{fig1}. 
The off-diagonal short range order through the molecules (matrix $D$) 
is able to destroy the long range coherence produced by the 
random distribution of $R$ (gas). We are in the $P_2$ phase. 
The middle of Fig.~\ref{fig3a} shows the spectral distribution for the lowest 
critical value $d_*=1$. The lower of Fig.~\ref{fig3a} shows the case with 
$d<d_*=1$. In the latter, the spectral 
distribution is always in the $P_1$ phase. The off-diagonal short range order 
is not strong enough to destroy the coherence at zero virtuality. In this 
regime chiral symmetry is always spontaneously broken.

We note that instanton models with an average instanton density of one 
instanton per fm$^4$, display spectral distributions that are close to 
the critical distribution with $\alpha$ small \cite{EARLY}. 
The instanton-to-molecule model for the finite temperature transition in QCD
\cite{SHURYAKMOL},
corresponds to the situation where $\alpha$ interpolates between 1 and 0,
discontinuously in the case of a first order transition or smoothly in the 
case of a second or higher order transition.

\vskip 0.5cm

{\bf 5.\,\,\,\,\,\,}
We have shown how the method of Blue's functions can be applied to a large 
class of chiral random matrix models. In doing so, we have unfolded two 
algebraic equations that allow  a generic classification of these spectra. 
Kalkreuter's spectral distribution for one flavor and Wilson fermions, 
as well as the massless Dirac operator in QCD on the cylinder at high 
temperature (one-Matsubara), have been shown to fall into the Cardano class 
(cubic equation). Wilson fermions for 2+1 flavors yield a spectrum that 
follows from a 
a linear superposition of Cardano's solutions. Finally, instanton models
with off-diagonal short range order fall into the Ferrari class 
(quartic equation).

In terms of Blue's functions, we have discussed the role played by 
the end-point singularities and shown how they could be used to assess
the critical parameters. We have suggested that the spectral density of states 
at the merging end-points (zero-virtuality for instance) could be used as a 
pertinent order parameter for monitoring phase changes in QCD inspired spectra.
This is particularly relevant in the massive case, where the quark condensate 
ceases to be (strictly speaking) a good order parameter. In doing so, we have 
unraveled a rich phase structure.

The method we have followed in this paper is very powerful and allows for a 
number of future investigations. In particular, it allows for a simple 
assessment of the thermal aspects of the 
QCD phase diagram for 2+1 flavors using chiral random matrix 
models. Detailed lattice simulations are by now available to allow for a
quantitative understanding in terms of the spectral distributions.
This issue will be discussed next.

\vglue 0.6cm
{\bf \noindent  Acknowledgements \hfil}
\vglue 0.4cm
This work was supported in part  by the US DOE grant DE-FG-88ER40388,
by the Polish Government Project (KBN) grant 2P03B19609 and by 
the Hungarian Research Foundation OTKA.

\vskip 1cm
\setlength{\baselineskip}{15pt}

\end{document}